**Superiority of mild interventions against COVID-19 on public health and economic measures**

*Author information*

Makoto Niwa (1)(2)(3) E-mail: gr0486se@ed.ritsumei.ac.jp

Yasushi Hara (3) E-mail: yasushi.hara@r.hit-u.ac.jp

Yusuke Matsuo (4) E-mail: yusuke-matsuo@retty.me

Hodaka Narita (4) E-mail: hodaka-narita@retty.me

Lim Yeongjoo (5) E-mail: lim40@fc.ritsumei.ac.jp

Shintaro Sengoku (6) E-mail: ssengoku@ifi.u-tokyo.ac.jp

Kota Kodama* (1)(3)(7) E-mail: kkodama@fc.ritsumei.ac.jp

(1) Graduate School of Technology Management, Ritsumeikan University, 2-150, Iwakura-cho, Ibaraki, Osaka 567-8570, Japan.

(2) Discovery Research Laboratories, Nippon Shinyaku Co., Ltd. Nishinosho-Monguchicho 14, Minami-ku, Kyoto 601-8550, Japan.

(3) TDB Center for Advanced Empirical Research on Enterprise and Economy, Faculty of Economics, Hitotsubashi University, 2-1 Naka, Kunitachi, Tokyo 186-8603, Japan.

(4) Retty, Inc.

(5) Research Organization of Open Innovation & Collaboration Research Center for

Innovation Management, Ritsumeikan University, 2-150, Iwakura-cho, Ibaraki, Osaka 567-8570, Japan.

(6) Life Style by Design Research Unit, Institute for Future Initiatives, the University of Tokyo, 7-3-1 Hongo Bunkyo-ku, Tokyo 113-0033, Japan.

(7) Center for Research and Education on Drug Discovery, The Graduate School of Pharmaceutical Sciences in Hokkaido University, Sapporo 060-0812, Japan

* Corresponding author

*Contributions*

M.N., and K.K. conceived of the study. M.N., Y.H. and K.K. designed the database. M.N. Y. N. H. N. and Y. H. collected the data. M.N. and Y. H. analysed the data. Y. L., S. S. and K. K. managed the project.

## Abstract

During the global spread of COVID-19, Japan has been among the top countries to maintain a relatively low number of infections, despite implementing limited institutional interventions. Using a Tokyo Metropolitan dataset, this study investigated how these limited intervention policies have affected public health and economic conditions in the COVID-19 context. A causal loop analysis suggested that there were risks to prematurely terminating such interventions. On the basis of this result and subsequent quantitative modelling, we found that the short-term effectiveness of a short-term pre-emptive stay-at-home request caused a resurgence in the number of positive cases, whereas an additional request provided a limited negative add-on effect for economic measures (e.g. the number of electronic word-of-mouth (eWOM) communications and restaurant visits). These findings suggest the superiority of a mild and continuous intervention as a long-term countermeasure under epidemic pressures when compared to strong intermittent interventions.

In the international context, social responses to COVID-19 have comprised various non-pharmaceutical interventions (NPIs) [1][2], including lockdowns. Although these measures may help prevent the spread of infection during the acute disease phase, many strong NPIs have resulted in obvious negative effects [3]. This highlights the need to focus on NPI optimisation, especially to achieve sustainable social responses against new infectious diseases.

In this study, we focused on Japan, which is an Organisation for Economic Co-operation and Development (OECD) country that has implemented various tiered countermeasures designed to minimise the economic impacts of COVID-19 [4]. In our analysis, we specifically focused on the dynamics of infection, NPIs, and social impacts. Despite the early outbreak of COVID-19 across many densely populated urban areas [5], Japan has kept its infection numbers low, particularly when compared to the numbers in other early-breaking countries (Figure 1). As for specifics, Japanese NPIs have discouraged the visitation of crowded locations [6] through so-called 'mild lockdowns', which refer to non-coercive stay-at-home requests [7]. Because of the low numbers seen across Japan, it is highly important to analyse the dynamics of these mild lockdown conditions as sustainable countermeasures prior to the subsequent sub-acute/chronic phase of COVID-19 control (e.g. mild lockdowns and/or new-normal lifestyles), when increased disease resistance may be achieved through widespread vaccination.

[Figure 1 here]

Previous studies have extensively investigated the negative effects of COVID-19 on different areas of business, especially in the service industry. For example, Anguera-Torrell et al. [8] examined the effects of COVID-19, economic stimulus, and government-sponsored loans on stock prices in the hotel industry, thus showing that COVID-19 cases presented negative effects, while loans showed positive effects. Meanwhile, other studies have focused on NPIs, including Yang et al. [9], whose investigation of the restaurant industry confirmed that stay-at-home orders led to lower demand; here, voucher programs were suggested as appropriate relief strategies. These businesses should also develop contingency plans in which transactions are targeted at takeout and delivery services, thereby minimising the need for human contact. In this regard, contactless and/or digital payment technologies can produce reinforcing effects. Another contactless example pertaining to sales and advertising is communication via electronic word-of-mouth (eWOM), which facilitates social distancing. Indeed, previous research has shown that eWOM communication positively impacts sales [10]. In this case, increased sales may be achieved through a concomitant rise in eWOM popularity during COVID-19. This makes it especially important to explore the dynamics of eWOM communication in the pandemic context.

Based on the information above, this study investigated both the dynamics of infection and the social impacts of NPIs in the Tokyo Metropolitan area, thus contributing to the literature on sustainable infection control in densely populated areas. From a technical perspective, we aimed to handle the potential of different responses to public health and economic issues found in the complex metropolitan dataset by implementing structured models. In this context, we investigated how several types of NPIs affected public health in general (based on the number of patients) and specifically within the restaurant industry (using metrics such as the number of customer visits and eWOM communications).

As a research strategy, we employed system dynamics, which is a powerful tool for investigating 'what-if' scenarios in complex situations. Initially, we conducted a causal loop analysis and employed a quantitative stock-flow model to investigate the quantitative aspects in consideration of important identified factors. We constructed disease and NPI effects models based on epidemic data obtained from the Tokyo Metropolitan area, specifically based on Japanese NPI practices from March 2020 to September 2020. Meanwhile, the data used to explore eWOM dynamics were collected from 2019 to 2020. We simulated the NPI effects based on different strategies and strengths, and also modelled and simulated the dynamics of both e-WOM communications and customer visits to restaurants.

## Results

*Causal loop diagram: Dynamics of infection*

 Though there is still debate on whether SARS-CoV-2 spreads via aerosols, the virus has resulted in what is known as the COVID-19 pandemic; in this regard, it is certain that SARS-CoV-2 spreads during close interpersonal contact [11]. In fact, the dynamics of COVID-19 transmission have successfully been described through the use of models showing that infections are spread during such exposure, especially among susceptible populations [12]. Further, asymptomatic carriers can still transmit the virus [13]. Here, transmission efficiency may be affected by a variety of factors, including travel [12] and population density [13][15][16][17]. In sum, the current literature shows that human-to-human contact is a determinant of transmission. As such, the course of infection (including the role of asymptomatic carriers) was featured in the causal loop.

*Causal loop diagram: Reviewing the basis of Japanese NPI*

 Regarding those implemented in 2020, Japanese NPIs can be categorised on the basis of need, including mild tiered interventions (raising sanitation awareness, physical distancing, encouraging remote work and staying home), focused NPIs in high-risk settings (suspending night services for bars and restaurants), and strong NPIs in more severe cases (strongly asking people to remain home, temporary business closures, according to administrative order) [18]. Milder NPIs were implemented during the early phase of the domestic outbreak in February 2020, at which time behavioural requests were issued, such as remembering to wash hands and covering the mouth when coughing [4]. More stringent requests were introduced in March 2020, when individuals were advised to avoid crowded locations [6]. Later on, stronger NPIs were implemented based on a declared state of emergency, with key factors including disease severity, uncontrollability, and the potential of overwhelming hospitals. Stricter measures were also implemented based on the particular severity of COVID-19 in serious cases. In this regard, the spread of the disease inevitably entered an uncontrollable phase due to the silent nature of transmission and moderate reproduction rate. This also resulted in a shortage of hospital beds. As of April 2020, only 12,500 beds were available for those with novel infectious diseases in Japan, accommodating approximately 10,000 patients at the time [13]. The surge of COVID-19 infections thus impelled the Japanese government to declare a state of emergency, at which time hospital capacities were increased [14]. There were also concerns about the possible need to accept additional patients during subsequent epidemic phases, thereby putting health systems in further danger. Ultimately, a second state of emergency was declared in January 2021 as Japan entered the next wave of the pandemic.

It is generally understood that NPIs may negatively affect the economy due to restricted mobility, especially during lockdowns [19]. In Japan's case, economic countermeasures were targeted at the travel industry, which was expected to benefit from anti-epidemic prioritisation in the NPI context during July 2020, when the number of patients in Tokyo (the most densely inhabited area) indicated the existence of a second viral wave [18]. This is evidence of how the economic sector may pressure the government to forego stronger NPI measures in times of crisis.

Based on the above, we considered that hospital bed shortages were key factors for inducing strong NPIs. On the other hand, we believed that concerns over economic stagnation would serve as negative feedback to stricter NPIs. As part of its public health policy, Japan's virus testing operations are focused on potential and identified clusters rather than mass testing; this is also a component of an active epidemiological survey [20]. In this regard, it would be adequate to consider that identifying virus carriers either by mass-oriented or focused approaches will help quarantine.

*Causal loop diagram: Behavioural effect on transmission and triggering behavioural change*

Among the various behavioural measures taken to avoid viral transmission, a system review has suggested the particular efficacy of physical distancing and the usage of face masks [11]. In Japan's case, previous studies have investigated behavioural changes during the very early phase (February 2020), thus highlighting the positive initial response triggered by mass-media reports of substantial infections incurred on a cruise ship, in which case the ministry's order to close schools was well-observed [21][22]. Further, an online survey conducted by the LINE Corporation and Ministry of Health and Welfare revealed behavioural changes in which individuals were more willing to avoid crowded locations as soon as Japan declared a state of emergency (April 2020) [23]. Based on these episodes, it would be adequate to consider that behavioural changes are induced via mass-media reports on disease threats and the need for strong NPIs.

*Causal loop diagram: People flow*

People flow refers to intercommunity human contact and can thus play a significant role in disease transmission, particularly in crowded areas. This is especially pertinent in Japan, which contains several cities with high population densities [5]. As such, we investigated the dynamics of people flow using data provided by Agoop, which is a big data company that collects location information related to people flow through smartphones [24].

For example, we examined the time course of people flow at Shinjuku Station in Tokyo, which is known for facilitating the largest number of commuters in Japan (as many as 250,000). People flow first decreased in March 2020, just after the ministry directed school closures. There was a drastic decrease in April 2020, when the stay-at-home request was issued. Figure 2 shows changes in the number of people at 18:00 on weekdays, which is considered the peak both around and within Shinjuku Station; this represents people flow during the typical commuting time in the Tokyo Metropolitan area. Even after the stay-at-home request was lifted, people flow was found to be at considerably lower levels when compared to pre-pandemic numbers. This shift to lower levels may indicate a general transition to a 'new-normal' lifestyle (e.g. remote work or behavioural changes in which people remain home or avoid visiting crowded locations due to long-term concerns about COVID-19). In this context, we constructed quantitative models to test the hypothesis that people flow in crowded places can affect disease transmission. This is described in more detail later in the manuscript.

[Figure 2 here]

*Causal loop diagram: Customer visits and eWOM*

To understand how NPIs affect the restaurant industry, we analysed data on actual customer visits and eWOM communications from Retty Inc., which runs an integrated web-based e-WOM and restaurant reservation service [25]. Both monthly metrics were taken from the year 2020 and normalised on the basis of the previous year. In short, results suggested that customer visits and eWOM (together with people flow) decreased because of COVID-19 in both March and April 2020, with further possible effects from the stay-at-home request issued in April 2020 (Figure 2). All three metrics were similarly affected by these events. Using monthly data from March to September 2020, the correlation coefficient between people flow and customer visits was 0.916, while that between people flow and eWOM was 0.879.

Further, the move in eWOM communications was slower than the shift in customer visits during the pandemic phase from March 2020 to May 2020 (includes the stay-at-home request). This suggests that, although eWOM follows customer visits, there is a delay in outcomes during crisis-driven circumstantial changes. Another hypothesis is that eWOM is resistant to pandemic crises because of the remote nature of the information transmission.

In general, we found that eWOM communications and customer visits were similarly affected by initial information and interventions related to COVID-19; indeed, full recovery was not observed for six months (October). This suggests that the pandemic has had long-term effects on public consciousness.

In sum, eWOM did not appear to affect customer visits in Tokyo based on the abovementioned time course dynamics. Rather, there was a tendency in which eWOM followed customer visits during the first pandemic wave. Although previous reports have shown that eWOM affects sales [10], most such findings are based on cross-sectional observations without clarifying specific conditional dynamics. This suggests the need for additional research once time-course data have been collected under various conditions.

*Constructing an integrated causal loop diagram*

Based on the information above, we constructed a causal loop diagram depicting the general situation for COVID-19 transmission in the context of Japanese NPIs, including the drawbacks (Figure 3A). Several feedback loops are present in the diagram. As for disease infection, COVID-19 transmission showed a reinforcing nature during the disease spreading phase (Figure 3B, upper); theoretically, herd immunity also appeared to decrease the transmission rate (Figure 3B, lower). On the other hand, the medical collapse loop illustrates the inadequacy of utilising herd immunity as a political strategy (i.e. allowing infections) (Figure 3C, lower), specifically showing that such practices are likely to result in uncontrollable transmission rates and insufficient medical treatments; in turn, this leads to a considerable increase in the death rate. Active epidemiological survey, which is a focused investigation on the circumstances of infection related to patients found, including virus testing of people contacted with patients [26] would partially decrease the transmission rate (Figure 3C, upper). As shown in Figure 3D, appropriate balancing dynamics and the strength of NPIs both decrease when the infection rate decreases. In contrast, the lower portion of Figure 3E shows negative feedback to strong NPIs based on economic stagnation, which can result in the premature termination of NPIs from a public health perspective.

Finally, the upper portion of Figure 3E shows the effects of eWOM. Previous research has clarified that increased eWOM is related to better sales [10]. In this study, we hypothesised there would be a short-term increase and decrease of eWOM mass in relation to sales.

[Figure 3 here]

While the nature of the 2020 pandemic has highlighted the need to prevent hospitals from becoming overwhelmed, there is still some reticence towards stronger NHIs, which can lead to economic stagnation. By contrast, delayed NHI implementation may worsen the COVID-19 mortality rate [16]. Practically speaking, a large number of hospital beds will be occupied for longer durations than normal, which puts serious pressure on hospital management. These conditions further emphasise the need to adequately time the implementation of appropriate NHIs. In this study, we investigated the issue by conducting a quantitative simulation, which is described in the following sections.

*Quantitatively modelling the disease transmission dynamics*

According to our model, disease transmission was accomplished via interpersonal contact, which was affected by both people flow and protective behaviour. Here, we assumed that people flow was affected by pandemic consciousness, stay-at-home requests, and new-normal lifestyle effects (Figure 4).

Short-term pandemic consciousness was considered a hypothetical psychological factor that could explain community resistance. In this study, it was introduced because transmission efficiency was found to vary over time, according to increases or decreases in reported patient numbers shown in previous research [13][27]. We hypothesised that media reports of increased cases would increase the level of pandemic consciousness, thus catalysing risk-evasive behaviours (e.g. not going out or wearing face masks and washing hands more frequently). Behavioural actions were parameterised as protective behaviour, then calibrated along the time course of real patient numbers; this effect was assumed to be reversible. The probability to engage in protective behaviour was set to 0.6 for the pandemic condition. This was derived on the basis of the results of a survey conducted by TDB-CAREE [28], which reported that about 60% of respondents believed more stringent measures were necessary in Tokyo as of June 2020. This subpopulation was therefore considered more likely to engage in protective behaviour. Meanwhile, new-normal lifestyle effects included the tendency to engage in remote work. This was handled as non-reversible based on the recognition that barriers to remote work included cyber security issues and employment rules.

As mentioned earlier, mass viral screening was not implemented in Japan, where virus testing was instead limited to symptomatic patients. During the early stages of the outbreak, typical symptoms were considered fever, fatigue, and/or shortness of breath. Our model was constructed in accordance with these conditions.

*Integrating a quantitative systems model across disease transmission, people flow, and the restaurant industry*

Our quantitative stock and flow model consisted of three components, including a disease transmission model, people flow and behaviour model, and effect on the restaurant industry model (Figure 4). First, the disease transmission model posited that virus carriers would transmit the virus to susceptible persons. We modelled symptomatic and non-symptomatic infections using data showing confirmed positive cases based on Japanese practices (i.e. virus testing was fundamentally limited to confirmative testing for symptomatic patients). Within the model, infections were considered dependent on the basic reproduction number [29], interpersonal contact, temperature [15], and the proportion of susceptible persons (i.e. non-immunised). On the other hand, we did not include virus mutations, possible vaccinations, or mass virus screening.

Second, the people flow and behaviour model posited that interpersonal contact related to disease transmission was affected by maximum people flow (i.e. locations for disease transmission) and personal behaviour (personal protective measures). Within the model, both elements were affected by psychological factors and short-term pandemic consciousness. Meanwhile, people flow was further dependent on extrinsic factors, such as stay-at-home requests and new-normal lifestyle effects, while behaviour was further affected by the distancing effect, behaviour guidance, and the thoroughness of protective behaviour.

Third, the restaurant industry model posited that customer visits to high-grade restaurants were dependent on the intention to dine out, but not necessarily dependent on people flow. As the interactions between customer visits and eWOM were unclear, they were modelled as independently affected by similar factors, such as stay-at-home requests, focused intervention effects, mid-term pandemic consciousness, long-term pandemic consciousness, and the psychological effect of school closures. Here, mid-term pandemic consciousness refers to a continuous mindset spanning months, particularly concerning the idea that individuals should voluntarily refrain from going out due to the risk of spreading disease. Next, long-term pandemic consciousness refers to a similar mindset that remains continuous for at least six months. This idea was introduced on the basis of the observation that customer visits and eWOM appeared to steadily react in contrast to fluctuating people flow. Finally, the psychological effect of school closures refers to both an initial recognition of the pandemic based on ministry-directed school closures [21][22] and a continued hypothetical psychological effect in which individuals avoid dining out as long as schools and other important educational facilities remain closed.

Though some part of the model structure was hypothetical, the quantitative aspects were calibrated based on real metrics, thus enabling useful simulations. In this regard, real

conditions were investigated through real data, which were also used as a basis of

comparison for the number of observed patients (confirmed positive viruses cases), people

flow (obtained via smartphone location information), number of visits, and eWOM

communication under realistic conditions.

[Figure 4 here]

*NPI pattern simulation*

We tested the effects of the four following interventional conditions: (A) realistic

conditions, in which one stay-at-home request lasting 1.5 months was issued in April 2020

(first pandemic wave). However, countermeasures against the second wave were limited to appeals for protective behaviour and remote work; (B) hypothetical stay-at-home request lasting 1 month was issued in July 2020, specifically as a countermeasure against the second pandemic wave; (C) hypothetical stay-at-home request lasting 1 month was issued in June, specifically as a pre-emptive countermeasure against a second pandemic wave; (D) an exhaustive intervention scenario, in which a first stay-at-home request lasting 2 months was issued in March 2020, specifically as a pre-emptive countermeasure against a first pandemic wave, with a second stay-at-home request lasting 2 months being issued in July 2020.

The number of patients (evaluated as confirmed positives) differed between scenarios (Figure 5A), with stay-at-home requests (especially when pre-emptively issued) lowering the number of patients. On the other hand, the more exhaustive intervention with stay-at-home request tended to result in more negative economic effects (Figure 5B), as represented by reduced customer visits. The outcome that was based on the direct effect size of the stay-at-home request was parameterised as a 10% decrease in customer visits and eWOM communication. More specifically, this parameterisation was based on the consideration that stay-at-home requests without closures (as actually directed in January 2021) do not constitute strong interventions. For reference, a previous study found a small add-on effect related to the lockdown condition [1].

There were two important findings. First, scenario (B) (pre-emptive stay-at-home request to counteract the second pandemic wave) effectively controlled the pandemic in the short-term context, with only small negative impacts to restaurant businesses; however, prematurely lifting the request would cause an explosive growth in the number of infections. Second, based on the current effect size of the employed factors, the economic effects of an additional lockdown were small, but the anti-pandemic effects were large. These findings indicate that a mild lockdown of substantial duration is an effective way to curtail the effects of the virus.

[Figure 5 here]

## Discussion

In this study, our causal loop analysis suggested that the fear of economic stagnation could urge societies to avoid or prematurely terminate strong NPIs, which are often necessary during viral pandemics. To avoid this, there must be ways to increase resilience among businesses that are vulnerable to new infectious diseases. This is particularly important for those operating in the restaurant, hotel, and travel industries. The appropriate use of mass media is also important, as news coverage is likely to influence behavioural change. Compared to the rates seen in other OECD countries, the relatively low number of COVID-19 patients found in Japan may have partially been the result of increased pandemic consciousness stemming from news coverage showing mass infections on cruise ships as well as the implementation of various administrative orders, including school closures and stay-at-home requests [21][22]. Regarding regional differences, it should be noted that there is room for research on biological ethnic differences such as pre-existing immunity [30], although we avoided the influence of factors which lead to severe outcomes by focusing on the number of positives.

Next, our quantitative model showed that pre-emptive and strong short-term interventions were efficient ways of controlling the spread of disease over short periods of time with minimum economic effects. However, these policies failed to control the spread of disease in the long-term context. The short-term efficiency of pre-emptive shorter lockdowns can be explained on the basis of the results of a preceding study, in which a cross-regional analysis [16] revealed that earlier lockdowns were more effective, while one-week delays in lockdown orders resulted in considerable disease spreading. In other words, while pre-emptive measures are effective, early lockdown terminations will still reboot the viral spread.

In sum, our simulations suggested that additional stay-at-home requests without school closures were effective, with only small negative effects. This was based on an assumption we built into the model; that is, initial fear and pandemic awareness triggered by mass media coverage and the ministry's order to close schools created an environment in which the public thoroughly followed recommended behaviours, such as engaging in personal protective measures. These results were also based on the recognition that it takes at least six months for customer visits and eWOM communications to recover from pandemic awareness. The analysis was further based on the assumption that only small negative impacts would be incurred through additional NPIs. Here, the efficacy of these additional interventions (typically implemented within six months of the initial intervention) constitutes an important practical finding. Combined with the usefulness of assessing the performance of various NPIs both alone and combined [1][2] as well as the flexibility of their operation [19], this finding should therefore facilitate the reduction of economic losses during NPIs.

As for practical insights, we found that the timeframes used for NPIs were highly important in the context of controlling new infectious diseases. Specifically, we recommend against the early termination of lockdowns, which is likely to result in the need for repeated implementation and lifting; this is not sensible. On the other hand, we recommend the continuation of mild sustained interventions, which are more likely to achieve long-term consistency in both public and economic health.

This study also had some limitations. One, the model did not explain neither within-family transmission nor the cluster effects and tended to underestimate patients as shown in Figure 5A. reliable effect sizes were not available for each NPI, in which case additional research should conduct time-course analyses under different conditions. Two, the studied time-course lasted from March 2020 to September 2020 in order to concentrate on the short-term effects of the pandemic and initial NPI. In this context, it is unlikely that we captured the long-term effects. Future studies should therefore focus on administrative activities that can support vulnerable industries. Nevertheless, our simulations employed a deductive model with hypothetical features that were supported through quantitative calibration, and were thus able to clarify complex societal systems and possible linkages between factors. This approach would also be useful in research targeted at the dynamics of other OECD countries, thereby pointing out areas of industrial similarity and/or social differences.

## Methods

*Data sources*

In this study, demographic data were obtained from a portal site containing official statistics relevant to Japan (e-Stat) [31]. The data were collected by the Ministry of Internal Affairs and Communications. On the other hand, we obtained the baseline frequency of dining out and statistical data pertaining to restaurant business from the Foodservice Industry Research Institute [32].

Next, e-WOM and customer visit metrics were gathered from a database compiled by Retty Inc., which runs an integrated web-based e-WOM and restaurant reservation service. People flow metrics were obtained from Agoop, which is a big data company that collects location information from smartphones [24]. Finally, data related to consumer sentiments were obtained through a monthly survey conducted by the TDB Center for Advanced Empirical Research on Enterprise and Economy (TDB-CAREE), Hitotsubashi University, Tokyo, Japan [28].

*System dynamics model linking disease spread, non-pharmaceutical countermeasures, and economic effects*

Both the causal loop and quantitative stock and flow models were built using Vensim PLE version 8.0.9 (Ventana Systems Inc.), while data summarisation was accomplished using Microsoft Excel 365 (Microsoft Corporation). The model was constructed through a bottom-up approach and calibrated according to real data. Initially, the full causal loop diagram was developed on the basis of reported events, preceding studies, and logical relationships. However, the model was reorganised into a quantitative stock and flow model. Appendix Table 1 shows the basis for our quantitative model parametrisation.

In the disease transmission model, transmission efficiency was described as the number of cases generated by one carrier and set to basic reproduction numbers under no affecting factors. Meanwhile, affecting factors were composed of people flow and behaviour components. The people flow component consisted of people flow in crowded places and a scaling factor that related basic reproduction numbers to real conditions. People flow that was responsible for transmission was represented by maximum people flow (persons per hour) on Wednesdays (middle of the week). The behaviour component was composed of several parameters, including distancing and protective behaviour, epidemic consciousness, fear of infection, and protective behaviour completeness; here, the effects were represented by relative values ranging from 0 to 1. As non-pharmaceutical countermeasures, stay-at-home requests and behaviour guidance were considered modulators of people flow and behaviour, respectively. Remote work practices were also thought to affect people flow.

The economic effect model described restaurant sustainability using eWOM. Here, the concept was that customer visits were affected by stay-at-home requests, epidemic consciousness, and positive eWOM. Restaurants with eWOM were assumed to be less susceptible to people flow and remote work. We also assumed that restaurants with positive eWOM had better sales, which we tested using the eWOM and customer visit metrics.

*Reality checking the quantitative model*

Figure Appendix 1 illustrates our comparison of the real and simulated data. We first confirmed that the quantitative model was of sufficiently good fit and that the approach was adequate. However, there were two notable limitations. First, the model tended to underestimate patients, as neither within-family transmission nor the cluster effects were modelled. However, the ability to describe historical disease spreading events implies that the model structure was useful for determining contributions and describing the dynamics of random transmission in crowded cities. Second, we did not adequately model the sharp decrease in customer visits found during April 2020. This may have been the result of a shock effect in which the population was facing a pandemic for the first time. As such,

additional research should target the influencing factors behind these dynamics.

Some of the NPI effects (possibly combined) could not be resolved because of the lack of sufficient time-course data with various conditions. The simulation was performed under the assumptive effect sizes of each NPI, while the effect sizes of school closures and stay-at-home requests were 20% and 10%, respectively. These NPI factors similarly affected people flow, customer visits, and eWOM communications. An NPI focused on suspending night services at restaurants was set to decrease only customer visits and eWOM communications (10% each). Regarding hypothetical factors, short-term epidemic consciousness was set to decrease people flow by 10%, while mid-term pandemic consciousness was set to decrease customer visits and eWOM by 20% each, and long-term pandemic consciousness was set to decrease customer visits and eWOM by 30% and 20%, respectively.

[Appendix Figure 1 here]

## Acknowledgements

This work was supported by Grants-in-Aid for Challenging Exploratory Research (grant numbers 20K20769, 26301022, and 23730336) and Ritsumeikan University With Corona Society Proposal Open Call Research Program. In addition, this work was supported by the FFJ/Air Liquide Fellowship. The author gratefully acknowledges the generous support and assistance of the Fondation France-Japon (FFJ) de Ecole des Hautes Etudes en Sciences Sociales (EHESS) and Air Liquide. The funding sources did not participate in study design, data collection, analysis, interpretation, report writing, or the decision to submit this article for publication. We appreciate the dataset provided by the TDB Center for Advanced Empirical Research on Enterprise and Economy (TDB-CAREE), which was used independently in this study, the human flow data provided by Agoop Corp. (https://www.agoop.co.jp/en/), and restaurant and eWON information provided by Retty, Inc. We would also like to thank all other open source and database developers.

## Conflicts of Interest

M.N. is an employee of Nippon Shinyaku Co., Ltd., a pharmaceutical company. The authors declare no other conflict of interest.

## References

[1] Haug, N. et al. Ranking the effectiveness of worldwide COVID-19 government interventions. *Nat. Hum. Behav.* **4,** (2020).

[2] Brauner, J. M. et al. Inferring the effectiveness of government interventions against COVID-19. *Science.* **371**(653) eabd9338 (2020) doi:10.1126/science.abd9338.

[3] Raboisson, D. & Lhermie, G. Living with COVID-19: A systemic and multi-criteria approach to enact evidence-based health policy. *Front. Public Heal.* **8,** 1–7 (2020).

[4] Headquarters for Novel Coronavirus Disease Control. Basic policies for novel coronavirus disease control. https://www.mhlw.go.jp/content/10200000/000603610.pdf (2020), viewed on 20 February 2021.

[5] Wendell Cox Consultancy. Demographia world urban areas, 16th annual edition. http://www.demographia.com/db-worldua.pdf (2020), viewed on 21 February 2021.

[6] Ministry of Health, Labour and Welfare. To prevent the outbreaks of novel coronavirus disease. https://www.mhlw.go.jp/content/10900000/000601720.pdf (in Japanese) (2020), viewed on 27 February 2021.

[7] Yamamoto, T., Uchiumi, C., Suzuki, N., Yoshimoto, J. & Murillo-Rodriguez, E. The psychological impact of 'mild lockdown' in Japan during the COVID-19 pandemic: A nationwide survey under a declared state of emergency. *Int. J. Environ. Res. Public Health*, **17**(24), 9382 (2020) doi: 10.3390/ijerph17249382.

[8] Anguera-Torrell, O., Aznar-Alarcón, J. P. & Vives-Perez, J. COVID-19: Hotel industry response to the pandemic evolution and to the public sector economic measures. *Tour. Recreat. Res.* **0,** 1–10 (2020).

[9] Yang, Y., Liu, H. & Chen, X. COVID-19 and restaurant demand: Early effects of the pandemic and stay-at-home orders. *Int. J. Contemp. Hosp. Manag.* **32,** 3809–3834 (2020).

[10] Rosario, A. The effect of electronic word of mouth on sales: A meta-analytic review of the effect of electronic word of mouth on sales: A meta-analytic review of platform, product, and metric factors. *J. Marketing Res.* **53**(3), 287-318 (2016) doi:10.1509/jmr.14.0380.

[11] Chu, D. K. et al. Physical distancing, face masks, and eye protection to prevent person-to-person transmission of SARS-CoV-2 and COVID-19: A systematic review and meta-analysis. *Lancet* **395,** 1973–1987 (2020).

[12] Kucharski, A. J. et al. Early dynamics of transmission and control of COVID-19: A mathematical modelling study. *Lancet Infect. Dis.* **20,** 553–558 (2020).

[13] Niwa, M., Hara, Y., Sengoku, S. & Kodama, K. Effectiveness of social measures against COVID-19 outbreaks in selected Japanese regions analyzed by system dynamic modeling. *Int. J. Environ. Res. Public Health* **17,** 1–12 (2020).

[14] Kato, S., Miyakuni, Y., Inoue, Y. & Yamaguchi, Y. Maximizing healthcare capacity in response to COVID-19 outbreak: Rapid expansion through education by health emergency and disaster experts. *Disaster Med. Public Health Prep.* 1–3 (2020) doi:10.1017/dmp.2020.264.

[15] Rubin, D. et al. Association of social Distancing, population density, and temperature with the instantaneous reproduction number of SARS-CoV-2 in counties across the United

States. *JAMA Netw. open* **3,** e2016099 (2020).

[16] Gerli, A. G. et al. COVID-19 mortality rates in the European Union, Switzerland, and the UK: Effect of timeliness, lockdown rigidity, and population density. *Minerva Med.* **111**(4), 308-314 (2020) doi:10.23736/S0026-4806.20.06702-6.

[17] Behnood, A., Mohammadi Golafshani, E. & Hosseini, S. M. Determinants of the infection rate of the COVID-19 in the U.S. using ANFIS and virus optimization algorithm (VOA). *Chaos, Solitons and Fractals* **139,** 110051 (2020).

[18] Anzai, A. & Nishiura, H. 'Go To Travel' campaign and travel-associated coronavirus disease 2019 Cases: A descriptive analysis. *J. Clin. Med.* **10**(3), 398 (2021).

[19] Rahman, M. A. et al. Data-driven dynamic clustering framework for mitigating the adverse economic impact of Covid-19 lockdown practices. *Sustain. Cities Soc.* **62,** 102372 (2020).

[20] The Diplomat. Japan's pragmatic approach to COVID-19 testing. https://thediplomat.com/2020/06/japans-pragmatic-approach-to-covid-19-testing/ , viewed on 20 February 2021.

[21] Machida, M. et al. Adoption of personal protective measures by ordinary citizens during the COVID-19 outbreak in Japan. *Int. J. Infect. Dis.* **94,** 139–144 (2020).

[22] Muto, K., Yamamoto, I., Nagasu, M., Tanaka, M. & Wada, K. Japanese citizens' behavioral changes and preparedness against COVID-19: An online survey during the early

phase of the pandemic. *PLoS One* **15,** 1–18 (2020).

[23] Ministry of Health, Labour and Welfare. Nationwide survey to act against novel coronavirus disease. https://www.mhlw.go.jp/stf/newpage_11109.html (2020) (in Japanese) viewed on 21 February 2021.

[24] Agoop corporation. About Agoop. https://www.agoop.co.jp/en (2021) viewed on 21 February 2021.

[25] Retty, Inc. Retty gourmet. https://retty.me/ (in Japanese) (2021) viewed on 21 February 2021.

[26] National Institute of Infectious Diseases. Active epidemiological investigation. https://www.niid.go.jp/niid/images/epi/corona/2019nCoV-02-200529.pdf (in Japanese) (2020) viewed on 21 February 2021.

[27] Niwa, M., Hara, Y., Lim, Y., Sengoku, S., Kodama, K. Optimizing social measures against COVID-19 using system dynamics modeling. ISPIM Connects Global 2020: Celebrating the World of Innovation - Virtual, 6-8 December 2020. Event Proceedings: LUT Scientific and Expertise Publications: ISBN 978-952-335-566-8.

[28] TDB Center for Advanced Empirical Research on Enterprise and Economy. RDB-CAREE Consumer psychological survey. https://www7.econ.hit-u.ac.jp/tdb-caree/survey/ (2020) (in Japanese) viewed on 21 February 2021.

[29] Xie, M. & Chen, Q. Insight into 2019 novel coronavirus — An updated interim review

and lessons from SARS-CoV and MERS-CoV. *Int. J. Infect. Dis.* **94,** 119–124 (2020).

[30] Le Bert, N. et al. SARS-CoV-2-specific T cell immunity in cases of COVID-19 and

SARS, and uninfected controls. Nature 584, 457–462 (2020).

[31] Statistics Bureau of Japan. e-Stat, portal site if official statistics of Japan.

https://www.e-stat.go.jp/ (2021) (in Japanese) viewed on 21 February 2021.

[32] Foodservice Industry Research Institute. About Foodservice Industry Research

Institute. http://www.anan-zaidan.or.jp/ (2021) (in Japanese) viewed on 21 February 2021.

[33] Park, M., Cook, A. R., Lim, J. T., Sun, Y. & Dickens, B. L. A systematic review of

COVID-19 epidemiology based on current evidence. *J. Clin. Med.* **9,** 967 (2020).

[34] Pascarella, G. et al. COVID-19 diagnosis and management: A comprehensive review. J.

*Intern. Med.* **288,** 192–206 (2020).

[35] Guan, W. et al. Clinical characteristics of coronavirus disease 2019 in China. *N. Engl. J.*

*Med.* **382,** 1708–1720 (2020).

**Figure Legends**

Figure 1. The number of cumulative positive COVID-19 cases in representative early-breaking OECD countries (United States, France, and Japan) and positive cases from the epicentre in China. Cases in the epicentre were reduced through strong interventions, while the proportion reached 10% in the United States (highest national proportion). The increase was more gradual in Japan.

Figure 2. People flow represented by the number present at 18:00 on weekdays, when there is typically a peak around and within Shinjuku Station (500 metre radius from the station centre) in the Tokyo Metropolitan area (left Y-axis); customer visits and eWOM mass (right Y-axis), recorded in 2020. Data were collected via smartphone location information, integrated web-based e-WOM, and restaurant reservation services.

Figure 3. Causal loop diagram for COVID-19 transmission reflecting Japanese NPIs. Solid arrows indicate relationships supported by policies, theories, or previous studies, while broken lines indicate hypothetical relationships. Restaurants were deemed representative businesses. Each panel shows A) the whole picture, B) reinforcing infection spreading loop and balancing herd immunity loop, C) balancing active epidemiological survey loop and reinforcing medical collapse loop, D) balancing people's behaviour loop and strong NPI loop, and E) reinforcing eWOM loop and balancing negative effects on strong NPI loop.

Figure 4. Integrated quantitative systems model across disease transmission, people flow, and the restaurant industry

Figure 5. (A) Simulation outcomes of confirmed positives, (B) customer visits, and (C) eWOM mass. Coloured bars on the upper side of the panel show duration of stay-at-home request (thick bars) and epidemic consciousness raised by information from 1st and 2nd wave outbreak (thin dotted bars) in each scenario (3rd wave outbreak was not taken into consideration)

Appendix Figure 1. Simulation outcomes for (A) people flow, (B) people dining out, and (C) eWOM mass compared to real metrics

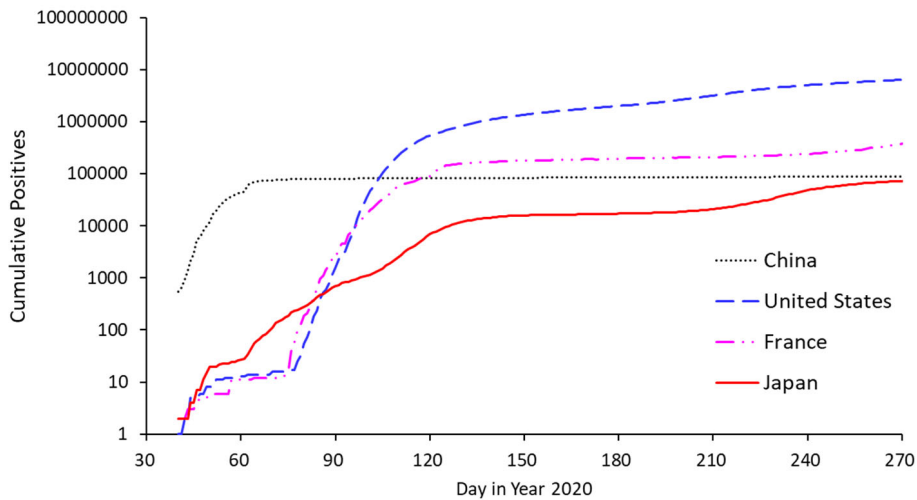

Figure 1. The number of cumulative positive COVID-19 cases in representative early-breaking OECD countries (United States, France, and Japan) and positive cases from the epicentre in China. Cases in the epicentre were reduced through strong interventions, while the proportion reached 10% in the United States (highest national proportion). The increase was more gradual in Japan.

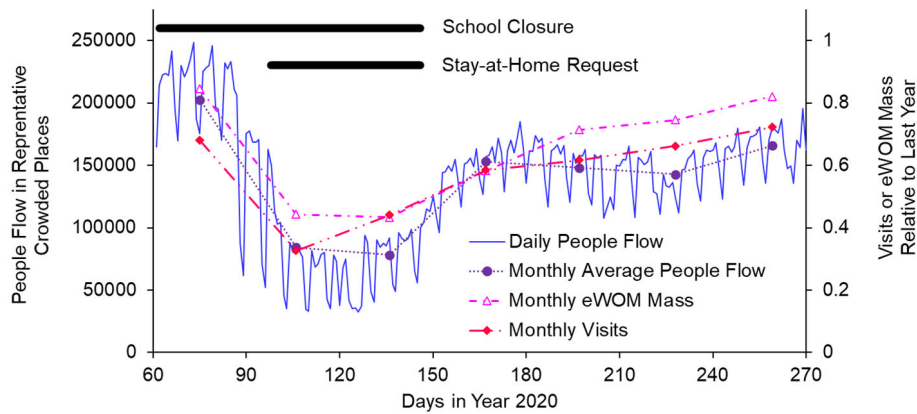

Figure 2. People flow represented by the number present at 18:00 on weekdays, when there is typically a peak around and within Shinjuku Station (500 metre radius from the station centre) in the Tokyo Metropolitan area (left Y-axis); customer visits and eWOM mass (right Y-axis), recorded in 2020. Data were collected via smartphone location information, integrated web-based e-WOM, and restaurant reservation services.

Figure 3. Causal loop diagram for COVID-19 transmission reflecting Japanese NPIs. Solid arrows indicate relationships supported by policies, theories, or previous studies, while broken lines indicate hypothetical relationships. Restaurants were deemed representative businesses. Each panel shows A) the whole picture, B) reinforcing infection spreading loop and balancing herd immunity loop, C) balancing active epidemiological survey loop and reinforcing medical collapse loop, D) balancing people's behaviour loop and strong NPI loop, and E) reinforcing eWOM loop and balancing negative effects on strong NPI loop.

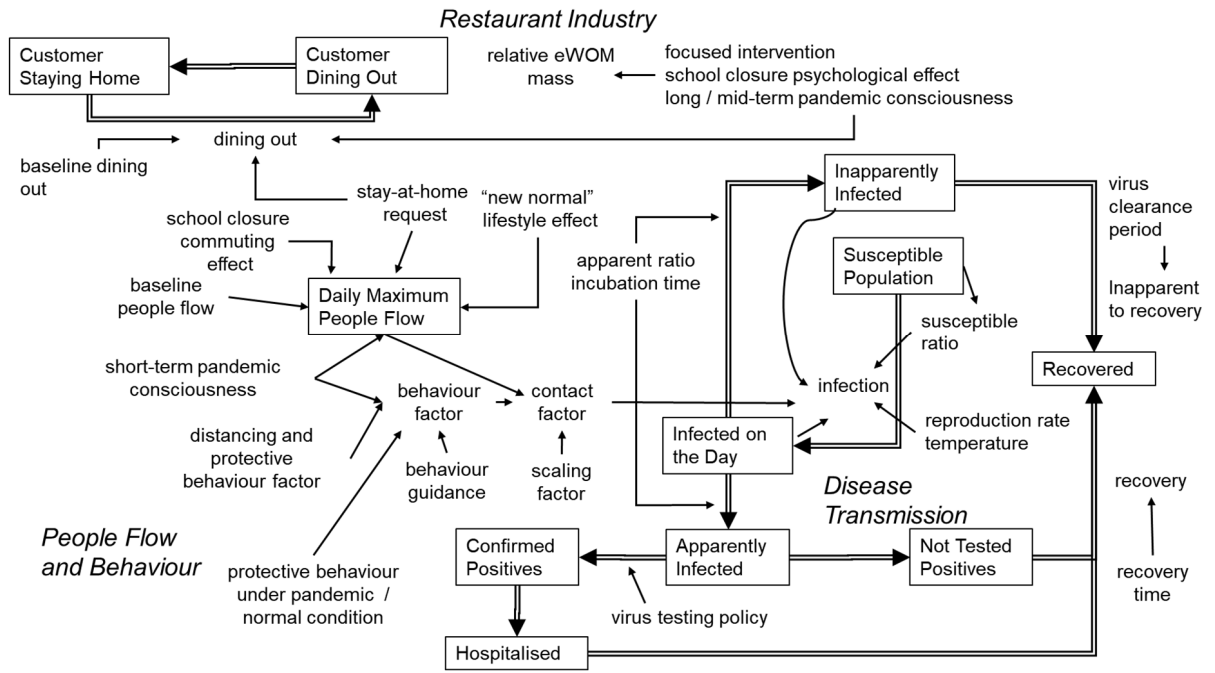

Figure 4. Integrated quantitative systems model across disease transmission, people flow, and the restaurant industry

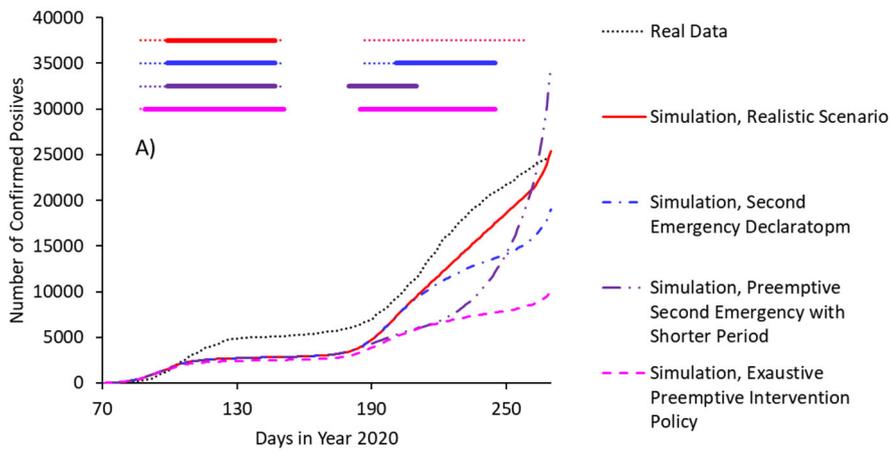

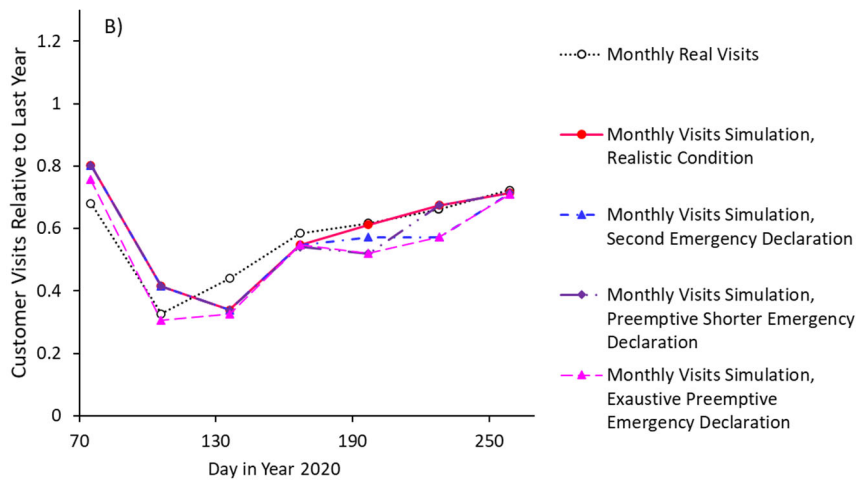

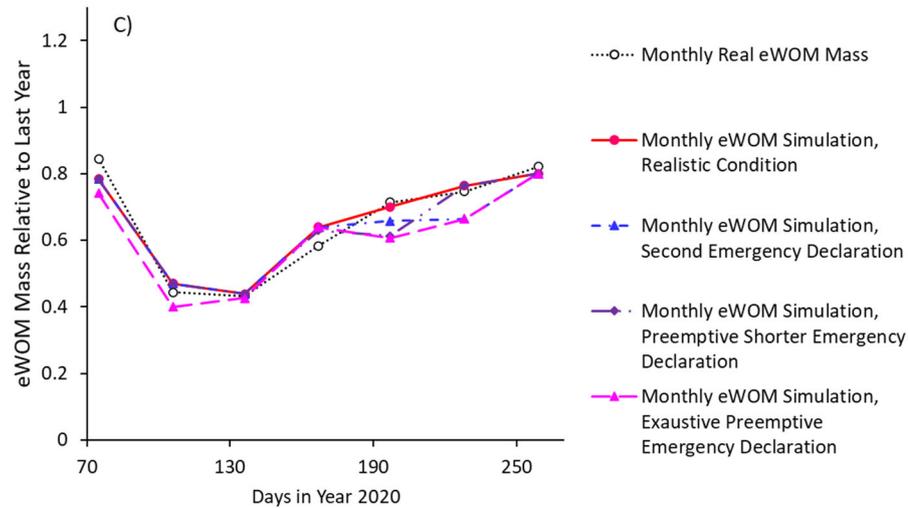

Figure 5. (A) Simulation outcomes of confirmed positives, (B) customer visits, and (C) eWOM mass. Coloured bars on the upper side of the panel show duration of stay-at-home

request (thick bars) and epidemic consciousness raised by information from 1st and 2nd wave outbreak (thin dotted bars) in each scenario (3rd wave outbreak was not taken into consideration)

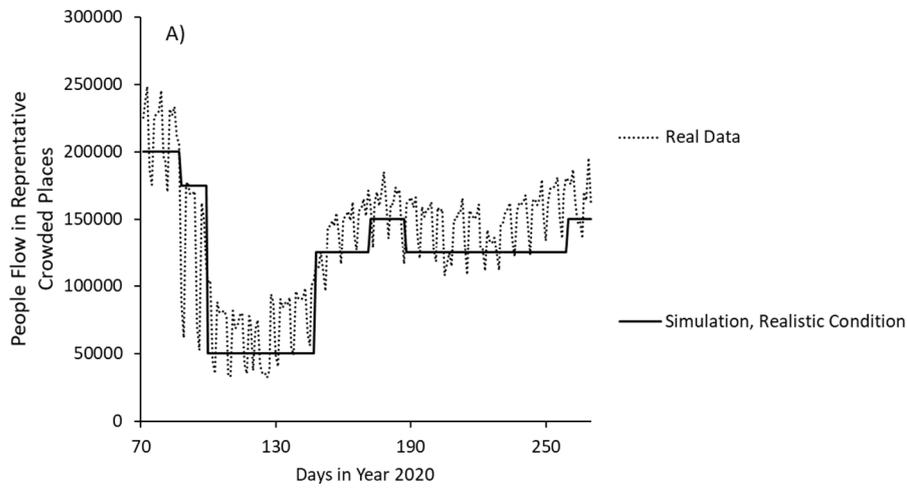

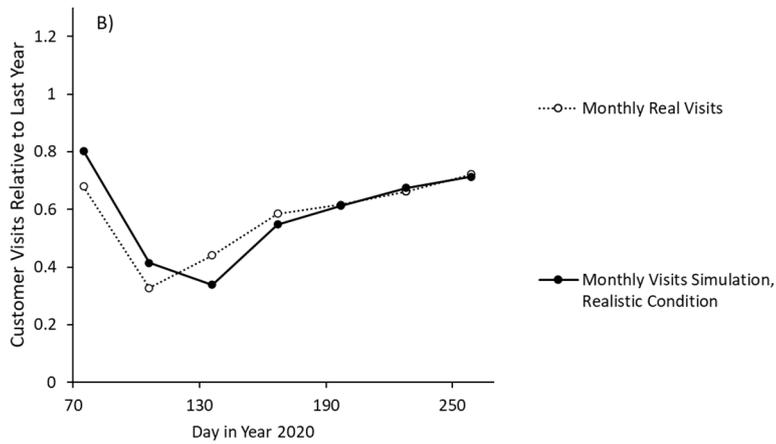

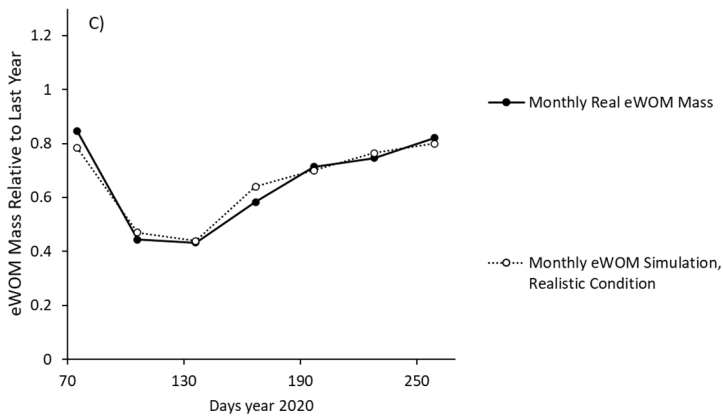

Appendix Figure 1. Simulation outcomes for (A) people flow, (B) people dining out, and (C) eWOM mass compared to real metrics

Appendix Table 1. Parametrisation of quantitative model

| Parameter | Meaning | Initial value | Equations | Comments |
|---|---|---|---|---|
| Disease Spreading Part | | | | |
| Susceptible | Susceptible (not immunised) people | 1.40 x 10$^7$ | -Infection | Whole Tokyo metropolitan population |
| Infected | People newly infected on a given day | 149 | infection-apparent Infection-inapparent infection | Initial value: calculated from actual confirmed positives after incubation period, apparent ratio, and testing policy |
| Apparent | People who newly appeared to be symptomatic | 60 | apparent Infection-not tested-virus testing symptomatic | Initial value: calculated from actual confirmed positives and testing policy |
| Inapparent | People inapparently infected and acting as virus carriers | 664 | inapparent infection-inapparent recovery | Initial value: calculated from actual confirmed positives, apparent ratio, and testing policy |
| Confirmed positives | People tested and confirmed as positive | 5 | virus testing symptomatic-hospitalisation | Initial value: actual confirmed positives |
| Not tested positives | People who are symptomatic and treated at home but not virus tested | 50 | not tested-not tested recovery | Initial value: calculated from actual confirmed positives and testing policy |
| Susceptible ratio | Effect of probability for a carrier to meet susceptible people | - | Susceptible/1.40 x 10$^7$ | - |

| Reproduction rate | Standard reproduction rate (2.9 / occasion) calibrated to a unit time (day) | 0.207 | - | 2.9: median of reported $R_0$ [33] |
|---|---|---|---|---|
| Temperature effect | Transmission affected by wet balb temperature [15] | 1.0 | 15 May 2020: 1.2<br>15 Jun 2020: 1.6<br>15 Jul 2020: 1.1<br>15 Sep 2020: 1.6 | Monthly average temperature was used |
| Apparent ratio | | 0.375 | - | [34] |
| Incubation time | | 5 | - | [34] |
| Apparent infection | | | Infected*apparent ratio/incubation time | - |
| Inapparent infection | | | Infected*(1-apparent ratio)/incubation time | - |
| Inapparent virus clearance period | | 8 | - | Tentative turnover 14 days minus infection day and incubation period |
| Testing policy | Initially, testing frequency was limited | 0.5 | Raised to 1.0 after 10 May 2020 | Half of the fevers are early onset [35] |
| People Flow and Behavior Part | | | | |
| Baseline People Flow | Baseline in representative crowded place | 250000 | - | 250,000 per representative traffic node |
| Daily Maximum People Flow | Maximum people flow in crowded places represented by station | 0 | Baseline People Flow*(1-0.2 x school closure commuting effect-0.1 x stay-at-home request -0.1 x short-term epidemic consciousness -0.4 x new normal lifestyle effect | - |

| Behaviour guidance | Exclusive guidance from Government | 0 | 15 Apr 2020: 1 | Recognition at this time point was suggested by LINE survey [23] |
|---|---|---|---|---|
| Distancing and protective behaviour facto | Reduced risk by individual protective behaviour (wearing masks, distancing) | 0.5 | - | [11] |
| Protective behaviour under epidemic condition | Probability for each person to act ideal protective behaviour | 0.6 | - | Estimated from the finding that about 60% of survey respondents thought more stringent measures were necessary in June 2020 [28] |
| Protective behaviour under normal condition | Probability for each person to act according to ideal protective behaviours | 0.3 | - | Half probability of epidemic condition |
| Restaurant Industry Part | | | | |
| Customer Staying home | | $1.07 \times 10^{7}$ | dining out to home-dining out | Initial value: population of Tokyo consisting of individuals 15 to 74 years of age |
| Customer Dining Out | | 0 | dining out-dining out to home | |
| Dining Out | Movement to dining out | 0 | baseline dining out x (1-0.2 x school-closure psychological effect-0.1 x stay-at-home request - | |

| | | | 0.1 x mid-term epidemic consciousness -0.1 x focused intervention effect -0.3 x long-term epidemic consciousness | |
|---|---|---|---|---|
| eWOM Mass | Relative daily eWOM mass | 0 | (1-0.2 x school-closure psychological effect -0.2 x long-term epidemic consciousness - 0.1 x stay-at-home request - 0.1 x focused intervention effect - 0.1 x mid-term epidemic consciousness) | |

Appendix Table 2. Parametrisation in each scenario

| Parameter | Realistic | Second Emergency | Pre-Emptive Shorter Emergency | Exhaustive Emergency |
|---|---|---|---|---|
| Short-term epidemic consciousness | Initial: 0<br>27 Mar 2020: 1<br>30 May 2020: 0<br>05 Jul 2020: 1<br>15 Sep 2020: 0 | Initial: 0<br>27 Mar 2020: 1<br>30 May 2020: 0<br>05 Jul 2020: 1<br>01 Sep 2020: 0 | Initial: 0<br>27 Mar 2020: 1<br>30 May 2020: 0<br>28 Jun 2020: 1<br>28 Jul 2020: 0 | Initial: 0<br>27 Mar 2020: 1<br>30 May 2020: 0<br>03 Jul 2020: 1<br>01 Sep 2020: 0 |
| Mid-term epidemic consciousness | Initial: 0<br>27 Mar 2020: 1<br>30 May 2020: 0 | Initial: 0<br>27 Mar 2020: 1<br>30 May 2020: 0 | Initial: 0<br>27 Mar 2020: 1<br>30 May 2020: 0 | Initial: 0<br>27 Mar 2020: 1<br>30 May 2020: 0 |
| Long-term epidemic consciousness | Initial: 0<br>27 Mar 2020: 1 | Initial: 0<br>27 Mar 2020: 1 | Initial: 0<br>27 Mar 2020: 1 | Initial: 0<br>27 Mar 2020: 1 |
| School closure, psychological effect | Initial: 1<br>26 May 2020: 0 | Initial: 1<br>26 May 2020: 0 | Initial: 1<br>26 May 2020: 0 | Initial: 1<br>26 May 2020: 0 |
| Stay-at-home request | Initial: 0<br>08 Apr 2020: 1<br>26 May 2020: 0 | Initial: 0<br>08 Apr 2020: 1<br>26 May 2020:0<br>19 Jul 2020: 1<br>01 Sep 2020 :0 | Initial: 0<br>08 Apr 2020: 1<br>26 May 2020: 0<br>28 Jun 2020: 1<br>28 Jul 2020: 0 | Initial: 0<br>29 Mar 2020: 1<br>30 May 2020: 0<br>03 Jun 2020: 1<br>01 Sep 2020: 0 |